\documentclass{ws-ijmpe}

\def\eps{\epsilon}

\def\inner#1#2{{\bm #1}\cdot{\bm #2}}
\def\Mc2{A A^\dagger -B B^\dagger+A B^\dagger-B A^\dagger}
\def\Md3{A A^\dagger -B B^\dagger-A B^\dagger+B A^\dagger}

\begin{document}

\markboth{Authors' Names}{Instructions for  
Typing Manuscripts (Paper's Title)}

\catchline{}{}{}{}{}

\title{A LAGRANGIAN FOR THE CHIRAL $(\frac12,0)\oplus
(0,\frac12)$ QUARTET NUCLEON 
RESONANCES
}

\author{\footnotesize V. DMITRA{\v S}INOVI{\' C}
}

\address{Vin\v ca Institute of Nuclear Sciences, lab 010,
P.O.Box 522, 11001 Beograd, Serbia
}

\author{ATUSHI HOSAKA}

\address{Research Center for
Nuclear Physics, Osaka University, Ibaraki 567-0047, Japan}

\author{KEITARO NAGATA}

\address{Department of Physics, Chung-Yuan Christian University, Chung-Li
320, Taiwan}
\maketitle

\begin{history}
\received{(received date)}
\revised{(revised date)}
\end{history}

\begin{abstract}
We study the nucleon and three $N^*$ resonances'
properties in an effective linear realization chiral $SU_{L}(2)
\times SU_{R}(2)$ and $U_{A}(1)$ symmetric Lagrangian. We place
the nucleon fields into the so-called ``naive" $(\frac12,0)\oplus
(0,\frac12)$ and ``mirror" $(0,\frac12)\oplus (\frac12,0)$
(fundamental) representations of $SU_{L}(2) \times SU_{R}(2)$, two
of each - distinguished by their $U_{A}(1)$ chiral properties, as
defined by an explicit construction of the nucleon interpolating
fields in terms of three quark (Dirac) fields. We construct the
most general one-meson-baryon chiral interaction Lagrangian
assuming various parities of these four nucleon fields. We show
that the observed masses of the four lowest lying nucleon states
can be well reproduced with the effective Lagrangian, after
spontaneous symmetry breakdown, without explicit breaking of
$U_{A}(1)$ symmetry. This does not mean that explicit $U_{A}(1)$
symmetry breaking does not occur in baryons, but rather that it
does not have a unique mass prediction signature that exists e.g.
in the case of spinless mesons. We also consider briefly the axial
couplings with chiral representation mixing. \end{abstract}

\section{Introduction}
\label{intro}

Chiral symmetry, as one of the symmetries of QCD, is a key to
understanding the dynamics of the strong interaction. In the real
world chiral symmetry is spontaneously broken, and plays a
dynamical role in various scattering processes involving the
Nambu-Goldstone bosons. Hadrons are then classified only according
to the residual vector symmetry and the full chiral symmetry can
be conveniently represented by the non-linear realization. Yet, as
pointed out by Weinberg~\cite{weinberg}, it still makes sense to
talk about irreducible representations of the complete chiral
symmetry group, and consider hadrons as mixtures of a (small)
number of such chiral multiplets. If so, one can use the chiral
symmetry as an algebraic symmetry that puts constraints on
physical observables, such as masses and coupling constants.
Furthermore, as chiral symmetry is restored at high temperature
and density, the change in hadron properties can be viewed as a
function of the change in the representation mixing. The present
paper is based on this point of view, where we extend some
previous work by Christos~\cite{Christos} for $N_f=2$, to include
chiral ``mirror" nucleon fields, with a view to ultimately
extending it to $N_f=3$, where some pioneering work has been done
in Refs.~\cite{Christos2,Zheng:1991pk,Chen:2008qv}. So, this
paper may be viewed, to some extent, as an intermediate
pedagogical step on the way to the full-blown problem. Yet, we
shall see that this simpler $N_f=2$ case contains most of the
features of the $N_f=3$ case, and that certain characteristic
problems become clearer without the algebraic complexity of the
$N_f=3$ case.

It is, therefore, important to determine the starting point here,
{\it viz.} the chiral multiplets of bare hadrons, in particular of
baryons, which are then mixed by the interactions to produce the
physical/dressed hadrons. Phenomenologically, the success of the
quark model implies the dominance of three valence quark component
in nucleon states \footnote{A reasonable reproduction of the
nucleon ground state properties in the quenched lattice QCD may
also be seen as a validation of this point of view.}. Our strategy
is therefore to choose nucleon chiral multiplets guided by the
nucleon fields written in terms of three quarks. Strictly
speaking, in the broken symmetry world the chiral structure of the
interpolating field is not identical to that of the physical state
that is coupled to the field; generally, the latter could be more
complicated than the former. In this sense, our choice of chiral
multiplets is perhaps the simplest one consistent with QCD.

The chiral multiplets associated with the lowest Fock space
components of the $I(J)=\frac12(\frac12)$ (nucleon) fields are
$(\frac12,0)\oplus (0,\frac12)$ and $(1,\frac12)\oplus
(\frac12,1)$. For higher Fock space components (e.g. pentaquarks,
septaquarks etc.) one finds chiral multiplets $(\frac32,1)\oplus
(1,\frac32)$, $(\frac52,2)\oplus (2,\frac52)$ etc., all of which
we shall ignore here. Indeed we shall not even include the
$(1,\frac12)\oplus (\frac12,1)$ multiplet, which appears only
with non-local three-quark interpolating fields. The
classification of local three-quark nucleon fields into chiral
multiplets has been recently worked out in Ref.~\cite{Nagata:2008zzc}.
Following up on this, we use the $U_{A}(1)$ and $SU_{L}(2) \times
SU_{R}(2)$ chiral (here and in what follows we shall refer to them
as the Abelian and the non-Abelian chiral symmetries,
respectively) transformation properties of the two independent
$J=\frac12$ local nucleon fields to write down and classify the
possible nucleon-meson interaction terms in the present paper.

Nucleon fields containing no derivatives are natural for the
even-parity ground state nucleons; in that case only the
$(\frac{1}{2},0) \oplus (0,\frac{1}{2})$ non-Abelian multiplet is
allowed by the Pauli principle \cite{Nagata:2008zzc}. It turns out, however,
that there are two such linearly independent fields. Their linear
combinations form different (``opposite") irreducible
representations of the $U_{A}(1)$ symmetry: one with the axial
baryon number $-1$ and another with axial baryon number
$+3$~\footnote{Jido et al~\cite{jido98} also noticed the curious
$U_{A}(1)$ transformation properties of one linear combination of
the nucleon fields, but did not mention that it was an Abelian
``mirror" assignment.}~\cite{jido98}. We shall use these
properties to classify the nucleon-meson interaction terms in the
present paper.

For odd-parity nucleon excited states, on the other hand, fields
with (at least) one derivative appear natural. Once we allow for
one space-time derivative to exist, we find nucleon fields with
chiral properties that are opposite, or complementary, i.e.
``mirror" to the non-derivative ``naive" ones \cite{lee72,detar},
e.g. the $(0,\frac{1}{2}) \oplus (\frac{1}{2},0)$. This fact
allows one to explicitly construct the interactions of the four
different nucleon fields with chiral mesons that can account for
both the masses and decays of the lowest-lying even- and
odd-parity nucleon resonances.

As a specific example, we choose four particular nucleon fields,
forming a $U_{A}(1)$ chiral nucleon quartet, that we identify with
the four lowest-lying nucleon resonances: the nucleon-Roper
even-parity pair and the $N^{*}(1535)$ and $N^{*}(1650)$
odd-parity resonances. We estimate the coupling strengths
from the nucleon masses. Our method applies equally well to any,
and not just the lowest-lying, $U_{A}(1)$ chiral nucleon quartet,
i.e., pair of nucleon parity doublets.

It turns out that there are many allowed one-meson-baryon
interaction terms, even in the limit of exact 
$SU_{L}(2) \times SU_{R}(2)$ and $U_{A}(1)$ 
chiral symmetries. This large latitude in
the theory stems from the existence of cubic-meson-field
interactions, first noticed by Christos \cite{Christos}, which
turn into one-meson-baryon interaction terms after spontaneous
breakdown of the chiral symmetry. The existence of
cubic-meson-field interactions, in turn, is a consequence of the
existence of nucleon fields with the Abelian axial charge $+3$. We
shall show that this fact implies that one does not need explicit
$U_A(1)$ symmetry breaking terms due to the $U_A(1)$ anomaly in
the nucleon sector to describe the nucleon masses and decays: they
can all be described without taking explicit $U_A(1)$ symmetry
breaking into account. This is in contrast with the case of purely
linear-meson-field interactions, which requires some explicit
$U_A(1)$ symmetry breaking terms in the nucleon sector to describe
the nucleon masses.

Consequently, the explicit $U_A(1)$ symmetry breaking, or its
restoration, do not need to have an impact on the nucleon
resonance spectrum. Note, however, that it is always possible to
incorporate the explicit $U_A(1)$ symmetry breaking effects by
including e.g. the 't Hooft interaction in the mesonic sector, or
even in the baryon sector of the Lagrangian without ``spoiling"
the pattern/ordering of baryon masses\footnote{their absolute
values may change, of course, but then they can be renormalized}.
This would not be the case if there were only one kind of nucleon
field, i.e., it is a consequence of the existence of two kinds of
nucleon fields with different axial baryon numbers.

It must be stated that a closely related study of the nucleon
ground state and one resonance has been done by Christos in
Ref.~\cite{Christos}, but without introducing the nucleon mirror
fields. Indeed, he studied only the two independent nucleon
non-derivative fields and related their masses to certain
$U_{A}(1)$ symmetry conserving interactions. One distinct
disadvantage of models without mirror nucleons is that
phenomenologically they can not describe the one-pion decays of
nucleon resonances at the tree level~\cite{jido}. This motivates
us to consider a class of models that includes mirror nucleons,
which task we complete in the present paper. There is another kind
of baryon chiral multiplet mixing: the $(\frac{1}{2},0) \oplus
(0,\frac{1}{2})$ and $(1,\frac{1}{2}) \oplus (1,\frac{1}{2})$
mixing. That is possible only when one allows derivatives in the
three-quark interpolating fields, and will be dealt with in a
separate paper \cite{NHD}.

This paper falls into five sections. After the Introduction, in
Sect. II we give a reminder of the basic facts regarding the
nucleon fields and their chiral transformation properties, as well
as derive the chiral properties of the new derivative fields. Then
in Sect. III we examine the nucleon-meson couplings and classify
the interaction terms according to their symmetry properties for
one particular pair of nucleon fields. In Sect. IV we calculate
some of the basic predictions of this effective interaction.
Finally, in Sect. V we summarize and draw our conclusions.

\section{Three-quark nucleon fields}

Firstly we examine the $SU_{L}(2) \times SU_{R}(2)$ and $U_{A}(1)$
transformation properties of various quark trilinear forms with
quantum numbers of the nucleon. This leads us to two pairs of
independent even and odd-parity nucleon resonances with
particularly simple $SU_{L}(2) \times SU_{R}(2)$ and $U_{A}(1)$
transformation properties.

Five non-vanishing, apparently different baryon local fields have
been explicitly constructed from three quark fields without
derivatives, such that the nucleon quantum numbers are properly
reproduced. Only two out of these five local trilinear fields, are
linearly independent, however \cite{Ioffe81}. Their propriety for
the job at hand can be tested in various ways: one way is by study
of their chiral symmetry transformation properties, another is by
directly applying them to the study of physical quantities, e.g.
by QCD sum rule or lattice QCD. In the QCD sum rule approach, it
was shown that one linear combination of the two independent
nucleon fields couples predominantly to the ground state nucleon,
while the one with orthogonal weights couples to the lowest-lying
odd-parity nucleon resonance $N^{*}$(1535)~
\cite{Jido:1996ia,jido98}.


\subsection{Even-parity nucleon fields}

For completeness' sake we show the five non-derivative objects
involving three quark fields
\begin{eqnarray}
N_{1}^+ &=&  \eps_{abc}({\tilde q}_{a} q_{b}) q_c,
\label{N_1}\\
N_{2}^+ &=&  \eps_{abc}({\tilde q}_{a}\gamma^5 q_{b}) \gamma^5
q_c,
\label{N_2}\\
N_{3}^+ &=&  \eps_{abc}({\tilde q}_{a}\gamma_{\mu} q_{b})
\gamma^{\mu} q_c,
\label{N_3}\\
N_{4}^+ &=&  \eps_{abc}({\tilde q}_{a} \gamma_{\mu} \gamma^5 {\vec
\tau} q_{b}) \cdot {\vec \tau} \gamma^{\mu} \gamma^5 q_c,
\label{N_4}\\
N_{5}^+ &=&  \eps_{abc}({\tilde q}_{a} \sigma_{\mu \nu} {\vec
\tau} q_{b}) \cdot {\vec \tau} \sigma^{\mu \nu} q_c , \
\label{N_5}
\end{eqnarray}
which are assigned as even-parity (as indicated by the superscript
$+$), spin 1/2, isospin 1/2 (``nucleon") fields. Here indices $a,
b$ and $c$ label the color of the three quarks, whereas the Pauli
matrices $\tau_i$ operate in the isospin space, and $q$ is the
light quark iso-doublet; we also define the ${\tilde q}$ to stand
for $q^{T} C \gamma_{5} i \tau_{2}$ in shorthand notation:
$${\tilde q} \equiv q^{T} C \gamma_{5} i \tau_{2},$$
where $C=i \gamma^2 \gamma_{0}$ and $\tau_{2}$ is the second Pauli
matrix in isospin space.

Ioffe~\cite{Ioffe81} used Fierz transformations to show that there
are two independent nucleon fields. Thus, one may use various
combinations of independent fields among Eqs. (\ref{N_1}) -
(\ref{N_5}) for the computation of two point correlation
functions, but the linearly independent choices are all equivalent, 
for instance, to Eqs. (\ref{N_1}) - (\ref{N_2}).


We shall try and systematize the nucleon fields according to both
their Abelian and non-Abelian chiral transformation properties and
show that this classification lends new meaning to certain
concepts introduced into the linear Gell-Mann--Levy sigma model
some time ago. As mentioned in the Introduction, Lee, DeTar and
Kunihiro, and Jido et al. \cite{lee72,detar,jido} used such a
model to calculate the even-odd parity nucleon mass difference as
well as the decay properties of the odd-parity resonances as a
function of their chiral transformation properties (``naive" or
``mirror").

In the QCD sum rule studies, it is well known that a typical
combination of a scalar-isoscalar nucleon field Eq.(\ref{N_1}) and
of the pseudoscalar-isoscalar one Eq. (\ref{N_2}) successfully
describes the properties of the nucleon. In addition, this choice
leads to one of the best known non-Abelian chiral transformation
properties, {\it viz.} the ``naive" non-Abelian one, while we find
a somewhat complicated Abelian chiral transformation law (that can
be reduced to a direct sum of a ``triple naive" and a ``mirror"
Abelian chiral transformations). This is the starting assumption
of this paper, and its consequences will be discussed extensively.

Let us consider the scalar-isoscalar $({\tilde q} q)$ and the
pseudoscalar-isoscalar $({\tilde q} \gamma^5 q)$ even-parity
nucleon fields
\begin{eqnarray}
 N_{1}^+ &=& \eps_{abc}({\tilde q}_{a} q_{b}) q_c,\\
 N_{2}^+ &=& \eps_{abc}({\tilde q}_{a}\gamma^5 q_{b}) \gamma^5 q_c .
 \label{e:nucleon+}
\end{eqnarray}
They transform according to the linear realization under the
non-Abelian chiral transformations:
\begin{eqnarray}
\delta_{5}^{\vec a} N_{1}^{+} &=& i \gamma_5 {\vec \tau} \cdot
{\vec a}
N_{1}^{+}, \\
\delta_{5}^{ \vec a} N_{2}^{+} &=& i \gamma_5 {\vec \tau} \cdot
{\vec a} N_{2}^{+}, \label{e:nuc+nAtrf}
\end{eqnarray}
whereas under the Abelian chiral transformations the rule is also
linear, but slightly more complicated as it mixes in the second
nucleon field:
\begin{eqnarray}
\delta_5 N_{1}^{+} &=& i a \gamma_5 (N_{1}^{+} + 2 N_{2}^{+}),
\\
\delta_5 N_{2}^{+} &=& i a \gamma_5 (N_{2}^{+} + 2 N_{1}^{+}).
\label{e:nuc+Atrf}
\end{eqnarray}
In other words they seem to form a two-dimensional representation
of the Abelian chiral symmetry $U_{L}(1) \times U_{R}(1)$, or an
$U_{A}(1)$ doublet. Of course, all irreducible representations of
an Abelian Lie group are one-dimensional. Therefore the ``chiral
doublet" two-dimensional representation of $U_{A}(1)$ furnished by
the fields $N_{1,2}$ and defined by Eqs. (\ref{e:nuc+Atrf1}) and
(\ref{e:nuc+Atrf}) must be a reducible one.


The symmetric and antisymmetric linear combinations of two
(identical parity) nucleon fields $N_{1,2}$ transform according to
the irreducible representation:
\begin{eqnarray}
N_{n}^{+} &=& \frac{1}{\sqrt{2}} (N_{1}^{+} + N_{2}^{+}), \\
N_{m}^{+} &=& \frac{1}{\sqrt{2}} (N_{1}^{+} - N_{2}^{+}).
\label{e:def}
\end{eqnarray}
Then their Abelian chiral transformation properties are
\begin{eqnarray}
\delta_5 N_{n}^{+} &=& 3 i a \gamma_5 N_{n}^{+},
\label{e:nuc+Atrf1} \\
\delta_5 N_{m}^{+} &=& - i a \gamma_5 N_{m}^{+} .
\label{e:nuc+Atrf2}
\end{eqnarray}
Note the factor 3 in front of the r.h.s. of Eq.
(\ref{e:nuc+Atrf1}), i.e., it is the ``triply-naive" Abelian axial
baryon charge transformation law, as it should be for an object
consisting of three quarks, and the negative sign in front of the
r.h.s. of Eq. (\ref{e:nuc+Atrf2}), as it should for an Abelian
``mirror" nucleon. The non-Abelian transformations remain
unchanged (``naive")
\begin{eqnarray}
\delta_{5}^{\vec a} N_{m,n}^{+} &=& i \gamma_5 {\vec \tau} \cdot
{\vec a} N_{m,n}^{+}. \label{e:nuc+Atrf4}
\end{eqnarray}

The triply-naive Abelian field does not have an $U_{A}(1)$
symmetric interaction that is linear in meson fields. The factor 3
on the right-hand side of the Abelian chiral transformation law
Eq. (\ref{e:nuc+Atrf1}) suggests, however, that the appropriate
power of meson fields should also be three, and, indeed, there are
two independent $U_A(1)$ invariants that are cubic in meson
fields, see \ref{app:cubic}. In other words, the $U_A(1)$
symmetry breaking is not intrinsic in the ``triple-naive" Abelian
nucleon structure.

Now, from the pair-wise nature of the nucleon fields under the
$U_A(1)$ transformations, it is natural to consider the nucleon
resonance states consisting of two parity doublets, i.e., of two
even-parity and two odd-parity nucleons. If all the resonances
belong to the naive non-Abelian representation, we can not avoid
the decoupling of off-diagonal $\pi NN$ interaction, as shown in
Ref.~\cite{jido}. Hence we need to find two fields with ``mirror"
non-Abelian chiral properties.

We have shown that it is not possible to construct a ``mirror"
non-Abelian nucleon from three quarks without derivatives
\cite{Nagata:2008zzc}. Without such a ``mirror" field, it is impossible to
have a pure ``naive-mirror mass term" that prevents the decoupling
of the pion interaction term. Therefore, let us try and see if
that can be done when one derivative is available.

\subsection{Odd-parity derivative nucleon fields}
\label{sect:nm-der}

We can construct nucleon fields with ``contravariant" chiral
transformations to those shown above by the replacement of
$\gamma_{\mu}$ with a derivative $i \partial_{\mu}$ (or a
covariant $i D_{\mu} =  i \partial_{\mu} + e A_{\mu}$ in QCD), for
example the following two derivative objects involving three quark
fields
\begin{eqnarray}
N_{1}^{\prime -} &=&  \eps_{abc} i \partial_{\mu} ({\tilde q}_{a}
q_{b})
\gamma^{\mu} \gamma^5 q_c,\\
N_{2}^{\prime -} &=&  \eps_{abc} i \partial_{\mu} ({\tilde q}_{a}
\gamma^5 q_{b}) \gamma^{\mu} q_c . \ \label{e:dTnucleon-}
\end{eqnarray}
They are odd-parity, spin 1/2 and isospin 1/2 fields, i.e. they
describe (some) nucleon resonances. A prime in the superscript
implies that the fields contain a derivative, and we show below
that therefore they have opposite, i.e., mirror, non-Abelian
chiral transformation properties to those of the corresponding
non-derivative fields.

Taking, once again, the symmetric and antisymmetric linear
combinations of two (identical parity) nucleon fields
$N_{1,2}^{\prime -}$ as the new canonical fields
\begin{eqnarray}
N_{m}^{\prime -} &=& \frac{1}{\sqrt{2}} (N_{1}^{\prime -} + N_{2}^{\prime -}), \\
N_{n}^{\prime -} &=& \frac{1}{\sqrt{2}} (N_{1}^{\prime -} -
N_{2}^{\prime -}) , \label{e:def2}
\end{eqnarray}
their Abelian chiral transformation properties read
\begin{eqnarray}
\delta_5 N_{m}^{\prime -} &=& - 3 i a \gamma_5 N_{m}^{\prime -},
\label{e:nuc+Atrf8} \\
\delta_5 N_{n}^{\prime -} &=& ~~ i a \gamma_5 N_{n}^{\prime -} ,
\label{e:nuc+Atrf9}
\end{eqnarray}
whereas the non-Abelian ones remain ``mirror"
\begin{eqnarray}
\delta_{5}^{\vec a} N_{m,n}^{\prime -} &=& - i \gamma_5 {\vec
\tau} \cdot {\vec a} N_{m,n}^{\prime -}. \label{e:nuc+Atrf10}
\end{eqnarray}
With these fields we are ready to construct ``naive-mirror"
interactions. We summarize the properties of the four fields in
Table.\ref{tab:prpB}

Thus, we have constructed four independent nucleon fields: two
fields with naive and two fields with mirror Abelian and
non-Abelian chiral transformation properties. In the present
paper, we identify these fields with the nucleon ground state
$N(940)$ and its resonances $N(1440)$, $N(1535)$ and $N(1650)$.

\begin{table}[tbh]
\begin{center}
\caption{The axial charges of the nucleon fields. Two generic
tentative assignments of physical states are shown as cases I and
II. In case I both even-parity fields are non-Abelian naive, in
case II the Roper is a mirror one.}
\begin{tabular}{ccccc}
\hline \hline
 & $U_A(1)$ & $SU_A(2)$ &(I) & (II) \\
 \hline
 $N_m$ & $-1$ & $+1$  & $N(940)$ & $N(940)$ \\
 $N_n$ & $+3$ & $+1$ & $N(1440)$ & $N(1535)$ \\
 $N_n^\prime$ & $+1$ & $-1$  & $N(1650)$ & $N(1650)$  \\
 $N_m^\prime$ & $-3$ & $-1$& $N(1535)$ & $N(1440)$ \\
 \hline
\end{tabular}
\label{tab:prpB}
\end{center}
\end{table}

\section{Nucleon-meson chiral interactions}
\label{sect:Lagrangian}

The previous studies \cite{lee72,detar,jido} developed a formalism
based on one pair of naive and mirror opposite-parity nucleon
fields. However, they did not make a reference to the $U_{A}(1)$
symmetry.  Our strategy is first to construct the $SU_{L}(2)
\times SU_{R}(2)$ chiral invariant interaction terms for
$N_{m,n}^{+}$ and $N_{m,n}^{\prime -}$ fields. These terms are
then classified according to their $U_{A}(1)$ symmetry. We shall see
that besides the usual linear (in meson fields) interactions there
are also quadratic and cubic ones. Christos \cite{Christos}, on
the other hand, has shown that there are two independent
three-meson-one-nucleon interactions for each parity doublet that
preserve both the Abelian and the non-Abelian chiral symmetry.
That makes altogether six terms: four diagonal ones in the two
doublets and two ``inter-doublet" ones, 
see \ref{app:cubic}. Furthermore, we can include quadratic terms that
are allowed by the non-Abelian mirror properties of the baryons.
We shall see that with so many $U_A(1)$ symmetry-conserving terms,
we do not need any $U_A(1)$ symmetry breaking terms to describe
this part of the nucleon mass spectrum, provided we use a complete
set of interactions (see Sect. \ref{sect:intB} below).

So far, we have considered identifications of $N_{m,n}^{+}$ with
positive parity states, while $N_{m,n}^{\prime -}$ with negative
parity states, as the operator containing a derivative may
naturally describe orbital excitations. In principle, we can also
consider the reversed case, however, where $N_{m,n}$ and
($N_{m,n}^{\prime }$) are identified with negative (positive)
parity states, respectively. In that case, the Lagrangians
constructed in this sections for the former identification are
transformed into those of the latter identification by multiplying
all the nucleon fields with a $\gamma_5$ matrix, e.g., $N^-
=\gamma_5 N^+$. We shall consider this possibility also in the
next section.

In the following, we consider two parity doublets
as follows; $\Psi = (N_m^{+},\; N_n^{\prime -})$ for the pair of
the single Abelian charge (single Abelian doublet), and $\Phi =
(N_n^{+},\; N_m^{\prime -})$ for that of the triple Abelian charge
(triple Abelian doublet). We emphasize that the two nucleons of
these pairs are in the "mirror" relations in both Abelian  and
non-Abelian chiral symmetries.

\subsection{``Single-Abelian" doublet}
\label{sect:intA}

First, we construct $U_{A}(1)$ symmetric Lagrangians from the
nucleon fields with single Abelian charge $N_{m}^{+}$ and
$N_{n}^{\prime  -}$. Since $N_{m}^{+}$ is the naive for the
non-Abelian chiral transformation, while $N_{n}^{\prime -}$ is the
mirror, then the $SU_L(2)\times SU_R(2)$ invariant interaction
terms up to first order of meson fields are as follows,
\begin{eqnarray*}
\bar{N}_{m}^{+} A N_{m}^{+},\; \bar{N}_{m}^{+} B N_{m}^{+},\;
\bar{N}_{n}^{\prime  -} A^\dagger N_{n}^{\prime  -},\;
\bar{N}_{n}^{\prime  -} B^\dagger N_{n}^{\prime  -},
\end{eqnarray*}
where we have introduced the meson fields,
\begin{eqnarray}
A &=& \sigma + i\gamma_5 \inner{\tau}{\pi},\\
B &=& \inner{\tau}{\sigma} + i\gamma_5\eta. \label{ABmeson}
\end{eqnarray}
In addition, chiral invariant mass term is also possible,
$\bar{N}_m^{+} \gamma_5 N_n^{\prime -} +h.c.$ Then, the $U_A(1)$
symmetric Lagrangian is formed by a suitable combination of the
interaction terms as
\begin{eqnarray}
{\cal L}_{\rm int}^{(1)} &=& - g_{1} \overline{N}_{m}^{+} (A - B)
N_{m}^{+} - g_{2} \overline{N}_{n}^{\prime -} (A^{\dagger} -
B^{\dagger}) N_{n}^{\prime -}
\nonumber \\
&& - m_{12} \big[\overline{N}_{m}^{+}\gamma_{5} N_{n}^{\prime -} +
{\rm h.c.}\big],
\nonumber \\
&=& - g_{1} \overline{N}_{m}^{+} \big[ \sigma -
\mbox{\boldmath$\tau$} \cdot \mbox{\boldmath$\sigma$} - i
\gamma_{5} (\eta -  \mbox{\boldmath$\tau$} \cdot
\mbox{\boldmath$\pi$}) \big] N_{m}^{+}
\nonumber \\
&& - g_{2} \overline{N}_{n}^{\prime -} \big[ \sigma -
\mbox{\boldmath$\tau$} \cdot \mbox{\boldmath$\sigma$} + i
\gamma_{5} (\eta - \mbox{\boldmath$\tau$} \cdot
\mbox{\boldmath$\pi$}) \big] N_{n}^{\prime -}
\nonumber \\
&& - m_{12} \big[\overline{N}_{m}^{+}\gamma_{5} N_{n}^{\prime -} +
{\rm h.c.}\big] \, . \label{e:mirrAb12}
\end{eqnarray}
Here the relative minus sign between $A$ and $B$ has been chosen
in order to preserve the $U_A(1)$ symmetry. Note that the $g_{1}$
and $g_{2}$ terms have different signs between $\sigma$ and $\pi$
fields that leads to differences in diagonalization and hence to
pionic nucleon decays. This shows that it is possible to break the
interlocking of mass and pion interaction matrices and thus allow
for pionic decays of nucleon resonances~\cite{jido}.

There are, moreover, three-meson interaction terms that preserve
both Abelian and non-Abelian chiral symmetries\footnote{We use
Christos' \cite{Christos} construction here.}
\begin{eqnarray}
{\cal L}_{\rm cubic~int}^{(1)} &=& - g_{3} f_{\pi}^{-2}
\overline{N}_{m}^{+} (A A^{\dagger} - B B^{\dagger}
+ A B^{\dagger} - B A{^\dagger})(A + B) N_{m}^{+}
\nonumber \\
&& - g_{4} f_{\pi}^{-2} \overline{N}_{n}^{\prime -} (A^{\dagger} +
B^{\dagger}) (A A^{\dagger} - B B^{\dagger}
- A B^{\dagger} + B A{^\dagger}) N_{n}^{\prime -},
\nonumber \\
&=& - g_{3} f_{\pi}^{-2} \overline{N}_{m}^{+}\big[ \sigma +
\mbox{\boldmath$\tau$} \cdot \mbox{\boldmath$\sigma$} + i
\gamma_{5} (\eta +  \mbox{\boldmath$\tau$} \cdot
\mbox{\boldmath$\pi$}) \big] N_{m}^{+} 
 \big( \sigma^{2} - \mbox{\boldmath$\sigma$}^{2} -
\eta^{2} + \mbox{\boldmath$\pi$}^{2} \big)
\nonumber \\
&& - 2 g_{3} f_{\pi}^{-2} \overline{N}_{m}^{+}\big[ i \gamma_{5}
(\sigma + \mbox{\boldmath$\tau$} \cdot \mbox{\boldmath$\sigma$}) -
(\eta + \mbox{\boldmath$\tau$} \cdot \mbox{\boldmath$\pi$}) \big]
N_{m}^{+} 
\big(\sigma \eta - \mbox{\boldmath$\sigma$} \cdot
\mbox{\boldmath$\pi$}\big)
\nonumber \\
&& - g_{4} f_{\pi}^{-2} \overline{N}_{n}^{\prime -}\big[ \sigma +
\mbox{\boldmath$\tau$} \cdot \mbox{\boldmath$\sigma$} - i
\gamma_{5} (\eta +  \mbox{\boldmath$\tau$} \cdot
\mbox{\boldmath$\pi$}) \big] N_{n}^{\prime -} 
\big( \sigma^{2} - \mbox{\boldmath$\sigma$}^{2} -
\eta^{2} + \mbox{\boldmath$\pi$}^{2} \big)
\nonumber \\
&& + 2 g_{4} f_{\pi}^{-2} \overline{N}_{n}^{\prime -}\big[ i
\gamma_{5} (\sigma + \mbox{\boldmath$\tau$} \cdot
\mbox{\boldmath$\sigma$}) + (\eta + \mbox{\boldmath$\tau$} \cdot
\mbox{\boldmath$\pi$}) \big] N_{n}^{\prime -} 
\big(\sigma \eta - \mbox{\boldmath$\sigma$} \cdot
\mbox{\boldmath$\pi$}\big). \label{e:mnA}
\end{eqnarray}
After spontaneous symmetry breaking, $\sigma \to f_{\pi} + s$,
however, this interaction leads to the same linearized
meson-nucleon interaction term as the linear interaction
Lagrangian Eq. (\ref{e:mirrAb12}) itself. Therefore for our
present purposes, ({\it viz.} establishing the nucleon masses and
two-body decay rates) the inclusion of this cubic term does not
generate any differences and for this reason we shall henceforth
drop this cubic interaction entirely.

As shown in Ref. \cite{jido,Christos} in some detail the linear
effective Lagrangian Eq. (\ref{e:mirrAb12}) is capable of
describing the $N(940)$ and the $N^{\prime -}_{n} = N^{*}(1535)$
or $N^{*}(1650)$ masses, while eliminating the single-pion decay
$N^{\prime -}_{n} \to \pi N(940)$ and retaining the decay
$N^{\prime -}_{n} \to \eta N(940)$.

\subsection{``Triple-Abelian" doublet}
\label{sect:intB}

Diagonal interactions for the nucleon fields with triple Abelian
charge, $N_n^+$ and $N_m^{\prime -}$,  is not as easily
constructed, since there is no term such that it preserves the
$U_{A}(1)$ symmetry and is linear in meson fields. As the desired
interaction must be cubic in meson fields, Christos'
\cite{Christos} construction has guided us in our quest.
\begin{eqnarray}
{\cal L}_{\rm cubic~int}^{(2)} &=& - g_{3} f_{\pi}^{-2}
\overline{N}_{n}^{+} (A A^{\dagger} - B B^{\dagger} - A
B^{\dagger} + B A{^\dagger})
 (A + B)N_{n}^{+}
\nonumber \\
&& - g_{4} f_{\pi}^{-2} \overline{N}_{m}^{\prime -} (A^{\dagger} +
B^{\dagger})
(A A^{\dagger} - B B^{\dagger} + A B^{\dagger} - B
A{^\dagger}) N_{m}^{\prime -}
\nonumber \\
&& - m_{34} \big[\overline{N}_{n}^{+} \gamma_5 N_{m}^{\prime -} +
{\rm h.c.}\big] ,
\nonumber \\
&=& - g_{3} f_{\pi}^{-2} \overline{N}_{n}^{+}\big[ \sigma +
\mbox{\boldmath$\tau$} \cdot \mbox{\boldmath$\sigma$} + i
\gamma_{5} (\eta +  \mbox{\boldmath$\tau$} \cdot
\mbox{\boldmath$\pi$}) \big] N_{n}^{+} 
\big( \sigma^{2} - \mbox{\boldmath$\sigma$}^{2} -
\eta^{2} + \mbox{\boldmath$\pi$}^{2} \big)
\nonumber \\
&& - 2 g_{3} f_{\pi}^{-2} \overline{N}_{n}^{+}\big[ i \gamma_{5}
(\sigma + \mbox{\boldmath$\tau$} \cdot \mbox{\boldmath$\sigma$}) -
(\eta + \mbox{\boldmath$\tau$} \cdot \mbox{\boldmath$\pi$}) \big]
N_{n}^{+} 
\big(\sigma \eta - \mbox{\boldmath$\sigma$} \cdot
\mbox{\boldmath$\pi$}\big)
\nonumber \\
&& - g_{4} f_{\pi}^{-2} \overline{N}_{m}^{\prime -}\big[ \sigma +
\mbox{\boldmath$\tau$} \cdot \mbox{\boldmath$\sigma$} - i
\gamma_{5} (\eta +  \mbox{\boldmath$\tau$} \cdot
\mbox{\boldmath$\pi$}) \big] N_{m}^{\prime -} 
\big( \sigma^{2} - \mbox{\boldmath$\sigma$}^{2} -
\eta^{2} + \mbox{\boldmath$\pi$}^{2} \big)
\nonumber \\
&& + 2 g_{4} f_{\pi}^{-2} \overline{N}_{m}^{\prime -}\big[ i
\gamma_{5} (\sigma + \mbox{\boldmath$\tau$} \cdot
\mbox{\boldmath$\sigma$}) + (\eta + \mbox{\boldmath$\tau$} \cdot
\mbox{\boldmath$\pi$}) \big] N_{m}^{\prime -} 
\big(\sigma \eta - \mbox{\boldmath$\sigma$} \cdot
\mbox{\boldmath$\pi$}\big)
\nonumber \\
&& - m_{34} \big[\overline{N}_{n}^{+} \gamma_5 N_{m}^{\prime -} +
{\rm h.c.}\big]. \label{e:3nnA}
\end{eqnarray}

There are, of course, many $U_{A}(1)$ symmetry breaking terms.
Their number, however, exceeds by far the number of available
observables, so we cannot hope to fix them from experiments alone.
More importantly, the $U_{A}(1)$ symmetry breaking terms are not
necessary to explain the mass spectrum, nor the decays within each
parity doublet. It is only the transitions/decays between members
of the two doublets that need to be added by hand.

Note, however, that {\it if we had (incorrectly) insisted on
interactions that are merely linear in meson fields, we would have
been led to a different conclusion {\it viz.} that the masses of
the ``triple Abelian" nucleon parity doublet are degenerate in the
good $U_{A}(1)$ symmetry limit}.

So, even though the mass splittings between the members of the
parity doublets are of the same order of magnitude as the explicit
$U_{A}(1)$ symmetry breaking scale, they may, in principle, be
unrelated to this symmetry breaking. Only a detailed model
calculation can distinguish between the symmetry-conserving and
-breaking contributions.

\subsection{Inter-doublet}
\label{sect:intC}

We look at the inter-doublet interaction terms for the pairs of
$N_m^+$ and $N_m^{\prime -}$, and $N_n^+$ and $N_n^{\prime -}$
\footnote{The naive-mirror pairings, e.g. $N_m^+$ and $N_n^{\prime
-}$, or $N_n^+$ and $N_m^{\prime -}$ are trivial, for either
parity combination.}. There are two $U_{A}(1)$ symmetry
invariants,
\begin{eqnarray}
{\cal L}_{\rm off-diag~int}^{(12)} &=& - g_{5} f_{\pi}^{-1}
\Big(\overline{N}_{m}^{+} (A A^{\dagger} - B B^{\dagger} 
+ A B^{\dagger} - B A{^\dagger})\gamma_5 N_{m}^{\prime -} +
{\rm h.c.} \Big)
\label{e:mirrAb2} 
\\
&& - g_{6} f_{\pi}^{-1} \Big(\overline{N}_{n}^{+} (A A^{\dagger}
- B B^{\dagger} 
- A B^{\dagger} + B A{^\dagger})\gamma_5 N_{n}^{\prime -} +
{\rm h.c.}\Big),
\nonumber \\
&=& - g_{5}f_{\pi}^{-1} \Big(\overline{N}_{m}^{+} \big[ \big(
\sigma^{2} - \mbox{\boldmath$\sigma$}^{2} - \eta^{2} +
\mbox{\boldmath$\pi$}^{2} \big)
- 2 i \gamma_{5} \big(\sigma \eta - \mbox{\boldmath$\sigma$}
\cdot \mbox{\boldmath$\pi$}\big) \big] \gamma_5 N_{m}^{\prime -} +
{\rm h.c.} \Big)
\nonumber \\
&& - g_{6} f_{\pi}^{-1} \Big(\overline{N}_{n}^{+} \big[
 \big( \sigma^{2} - \mbox{\boldmath$\sigma$}^{2}
- \eta^{2} + \mbox{\boldmath$\pi$}^{2} \big)
+ 2 i \gamma_{5} \big(\sigma \eta - \mbox{\boldmath$\sigma$}
\cdot \mbox{\boldmath$\pi$}\big) \big]\gamma_5 N_{n}^{\prime -}  +
{\rm h.c.} \Big) \, .
\nonumber 
\end{eqnarray}
There are no other (linear or cubic) meson interactions that
maintain both Abelian and non-Abelian chiral symmetries and
connect these two kinds of fields.

\subsection{Intra-doublet}

Next we consider the mixing between two non-Abelian-identical
members of the two doublets: e.g. between $N_{m}^+$ and $N_{n}^+$,
or between $N_{m}^{\prime -}$ and $N_{n}^{\prime -}$. As the
pion-nucleon interactions induced by these terms do not survive
the mass diagonalization, they are of limited practical use.
Because of this reason, we do not consider this type of
interactions in the following phenomenological study. Here,
however, we list them for the sake of completeness. There are two
terms with one meson field, that are given as
\begin{eqnarray}
{\cal L} &=& g_7 \left[\bar{N}_n^ +(A+B) N_m^+ +h.c.\right]
\nonumber \\
&+& g_8 \left[\bar{N}_n^{\prime -} (A^\dagger + B^\dagger)
N_m^{\prime -} + h.c.\right].
\end{eqnarray}
There are also two $U_A(1)$ invariant terms with three meson
fields, which are given as
\begin{eqnarray}
{\cal L}&=&g_9\left[\bar{N}_{n}^+ ( \Md3 )(A-B) N_{m}+h.c.\right]
\nonumber \\
+&g_{10}& \left[\bar{N}_{n}^-(A^\dagger-B^\dagger) (\Mc2)
N_{m}^-+h.c.\right].
\end{eqnarray}


\subsection{$U_{A}(1)$ symmetry breaking terms}

Finally and only for completeness' sake we consider the 
$U_{A}(1)$ symmetry breaking terms. Note that by changing the relative
sign of the $A$ and $B$ in all of the previous terms we break the
$U_{A}(1)$ symmetry while keeping the non-Abelian chiral
symmetry intact. Thus, for each $U_{A}(1)$ symmetry conserving
term there is (at least) one symmetry breaking term, and for cubic
interactions more than one, as one can change the relative sign in
several places. Thus, we see that for each $U_{A}(1)$ symmetry
conserving term there is a symmetry breaking one, and that the two
are indistinguishable with regard to their effects on the baryon
masses and pion interactions. The ``realistic" couplings contain
both kinds of terms, of course.

\subsection{Linearized nucleon-meson chiral interactions}
\label{s:intLin}

Upon taking into account the spontaneous symmetry breaking $\sigma
\to f_{\pi} + s$, we find the linearized forms of the linear,
quadratic and cubic interactions
\begin{eqnarray}
{\cal L}_{\rm int}^{(1)} &=& - g_{1} \overline{N}_{m}^{+}
\big[f_{\pi} + s - \mbox{\boldmath$\tau$} \cdot
\mbox{\boldmath$\sigma$} - i \gamma_{5} (\eta -
\mbox{\boldmath$\tau$} \cdot \mbox{\boldmath$\pi$}) \big]
N_{m}^{+}
\nonumber \\
&& - g_{2} \overline{N}_{n}^{\prime -} \big[ f_{\pi} + s -
\mbox{\boldmath$\tau$} \cdot \mbox{\boldmath$\sigma$} + i
\gamma_{5} (\eta - \mbox{\boldmath$\tau$} \cdot
\mbox{\boldmath$\pi$}) \big] N_{n}^{\prime -}
\nonumber \\
&& - m_{12} \big[\overline{N}_{m}^{+}\gamma_{5} N_{n}^{\prime -} +
{\rm h.c.}\big] \label{e:mirrAb12Lin} ,
\end{eqnarray}
and the linearized cubic term becomes
\begin{eqnarray}
{\cal L}_{\rm lin~cub~int}^{(1)} &=& - g_{3}
\overline{N}_{m}^{+}\big[f_{\pi} + 3 s + \mbox{\boldmath$\tau$}
\cdot \mbox{\boldmath$\sigma$} + i \gamma_{5} (3 \eta +
\mbox{\boldmath$\tau$} \cdot \mbox{\boldmath$\pi$}) \big]
N_{m}^{+}
\nonumber \\
&-& g_{4} \overline{N}_{n}^{\prime -}\big[f_{\pi} + 3 s +
\mbox{\boldmath$\tau$} \cdot \mbox{\boldmath$\sigma$} - i
\gamma_{5} (3 \eta +  \mbox{\boldmath$\tau$} \cdot
\mbox{\boldmath$\pi$}) \big] N_{n}^{\prime -}.
\label{e:cubmnA12Lin}
\end{eqnarray}
Note that Eq. (\ref{e:mirrAb12Lin}) and Eq. (\ref{e:cubmnA12Lin})
have the same shape as far as the mass and ${\vec \pi}$ terms are
concerned; the difference shows up only in the signs and sizes of
the $s$ and $\eta$ interactions, which can be three times
stronger, and of opposite sign. This means that we can put these
two interactions together and call the effective interaction
couplings $g_{1}, g_{2}$. For this reason we have used $g_{3},
g_{4}$ to denote the second (``triple-Abelian") doublet couplings
and thus reduce the nomenclature clutter.
\begin{eqnarray}
{\cal L}_{\rm lin~cub~int}^{(2)} &=& - g_{3}
\overline{N}_{n}^{+}\big[f_{\pi} + 3 s + \mbox{\boldmath$\tau$}
\cdot \mbox{\boldmath$\sigma$} + i \gamma_{5} (3 \eta +
\mbox{\boldmath$\tau$} \cdot \mbox{\boldmath$\pi$}) \big]
N_{n}^{+}
\nonumber \\
&& - g_{4} \overline{N}_{m}^{\prime -}\big[f_{\pi} + 3 s +
\mbox{\boldmath$\tau$} \cdot \mbox{\boldmath$\sigma$} - i
\gamma_{5} (3 \eta +  \mbox{\boldmath$\tau$} \cdot
\mbox{\boldmath$\pi$}) \big] N_{m}^{\prime -}
\nonumber \\
&& - m_{34} \big[\overline{N}_{n}^{+} \gamma_5 N_{m}^{\prime -} +
{\rm h.c.}\big]. \label{e:3nnAlin}
\end{eqnarray}
Finally, the linearized inter-doublet term is
\begin{eqnarray}
{\cal L}_{\rm off-diag~int}^{(12)} &=&
- g_{5} \Big(\overline{N}_{m}^{\prime -} \big[
f_{\pi}+ 2 s - 2 i \gamma_{5} \eta 
\big] N_{m}^{+}  + {\rm h.c.} \Big)
\nonumber \\
&&
- g_{6} \Big(\overline{N}_{n}^{+} \big[ 
f_{\pi} + 2 s + 2 i \gamma_{5} \eta \big] N_{n}^{\prime -}  + {\rm
h.c.} \Big). \label{e:3nnA2lin}
\end{eqnarray}
As anticipated, the parity may be reversed, and $N_m^\prime$ may
be assigned to be a positive parity field, for instance, the Roper
resonance. Then the quadratic $\pi$ interaction brings about the
necessary $R \to \pi \pi N$ decay strength in the even-parity
sector. The single-pion decay $R \to \pi N$ comes about due to the
``diagonal" interactions (within each parity doublet). Still the
odd-parity resonances $N^{*}(1535), N^{*}(1650)$ decays $N^{\prime
-}_{m} \to \pi N$ and $N^{-}_{m} \to \pi R$ are forbidden in the
good $U_{A}(1)$ symmetry limit, which can easily be corrected by
including (many different) $U_{A}(1)$ symmetry braking terms. This
mechanism for Roper decay does not exist if we assume that the
non-Abelian mirror field in the second doublet has odd parity.

\subsection{Choosing the members of parity doublets}
\label{s:members}

We have constructed the effective Lagrangian with the four baryons
$N_n,\; N_m, N_n^\prime$ and $N_m^\prime$. Now we proceed to the
assignment of these fields with the physical nucleon resonances,
which can be done by solving the diagonalization of the $4\times
4$ mass matrix. Of course, a complete exact diagonalization ought
to lead to the same (unique) solution, no matter what starting
point one adopts. That statement, however, holds only in the
idealized world in which all decays are kinematically allowed and
have been measured. Needless to say, we do not live in such a
world, so we must employ various tactics.

In the present paper, we rather employ a simple method with the
use of some insight from the lattice QCD and/or QCD sum rules,
which tell us that some observed baryons are dominated by
particular types of the baryon fields.

We begin with the classification of the four baryons into two
doublets $\Psi = (N_m, N_n^\prime)$ and $\Phi = (N_n, N_m^\prime)$
or their admixtures, with actual resonances {\it viz.} $N(940)$,
$R(1440)$ $N^{*}(1535)$ and $N^{*}(1650)$. Though having a larger
number of variations, we consider  two essentially different
scenarios.

As stated in the Introduction, a substantial body of QCD sum rule
evidence is pointing towards $N(940)$ as being the ``Ioffe
current" operator $N_{m}^{+}$. Together with the lowest
negative-parity nucleon $N(1535)$ as the partner in the parity
doublet, we have $\Psi = (N_m^+, N_n^{\prime -}) = (N(940),
N(1535))$ and $\Phi = (N_n^+, N_m^{\prime -}) = (N(1440),
N(1650))$. This is {\bf Scenario I}.

In another choice, we attempt to identify the negative parity
state $N(1535)$ with the non-derivative field $N_n^-$, as the QCD
sum rule implies substantial strength of the coupling between the
ground state. Hence we have $\Psi = (N_m^+, N_n^{\prime -}) =
(N(940), N(1650))$ and $\Phi = (N_n^-, N_m^{\prime +}) = (N(1535),
N(1440))$. This is {\bf Scenario II}.

This way of assigning the fields to states, rather than blind
solving the full $4\times 4$ mass matrices, may give us some
insight into the physical nature of the potential solution(s).
Now, we shall attempt to estimate the free parameters in each
case, so as to determine viability of either scenario.

\section{Results}
\label{sect:Results}

\subsection{Masses}
\label{s:mass}

Chiral symmetry is spontaneously broken through the
``condensation" of the sigma field $\sigma \to \sigma_0 = \langle
\sigma \rangle_{0} = f_{\pi}$, which leads to the dynamical
generation of baryon masses, as can be seen from the linearized
chiral invariant interaction Lagrangians Eqs.
(\ref{e:mirrAb12Lin})-(\ref{e:3nnA2lin}). Each of the two parity
doublets separately obeys the mass formulas given in Ref.
\cite{jido} for the ``mirror" case, provided there is no
interaction between the two doublets. We shall assume this at
first as the zeroth approximation, just to get a qualitative feel
for the results we may expect. There is additional mixing in the
four-nucleon mass/interaction matrix, however, due to the
``inter-doublet" interaction Eq.(\ref{e:mirrAb2}).

\begin{itemize}

\item {\bf Scenario I}

The nucleon mass matrix is already in a simple block-diagonal form
when the nucleon fields form the following 1$\times$4 row/column
``vector": $(\Psi,\Phi) = (N_{m}^{+}, N_{n}^{\prime -},N_{n}^{+},
N_{m}^{\prime -})$ $\to (N_{m}^{+}, \gamma_5 N_{n}^{\prime -},
N_{n}^{+}, \gamma_5 N_{m}^{\prime -})$
\begin{eqnarray}
\lim_{\rm U_A(1) symm.} M &=& \left(\begin{array}{cccc}
g_{1} f_{\pi} & m_{12} \gamma_{5} & 0 & g_{5} f_{\pi} \gamma_{5} \\
m_{12} \gamma_{5} & g_{2} f_{\pi} & g_{6} f_{\pi} \gamma_{5} & 0 \\
0 & g_{6} f_{\pi} \gamma_{5} & g_{3} f_{\pi} & m_{34} \gamma_{5} \\
g_{5} f_{\pi} \gamma_{5} & 0 & m_{34} \gamma_{5} & g_{4} f_{\pi} \\
\end{array}\right) .
\label{e:massM4} \
\end{eqnarray}
Note that only the parity-changing interaction $g_{5,6} \neq 0$
mixes these two new equal parity doublets. Manifestly we may
divide our analysis into two parts: one with, and another without
parity-flipping coupling.

First note that upon redefinition of odd-parity fields with a
$\gamma_5$, as in $N_{i}^{\prime} = \gamma_5 N_{i}^{-}$, where
$i=1,2$, the masses of the redefined fields pick up a minus sign.
This means that two of the mass eigenvalues will be negative.
Proper mass sign is restored at the end of the calculation when
one reverts back to odd parity fields, this time to diagonalized
ones, however.

The off-diagonal parity-non-changing coupling terms ($g_{5,6}$) in
the mass matrix do not appear to improve our ability to fit this
spectrum: rather they only seem to complicate the fitting
procedure. We shall set them equal to zero at first and use them
only later as they become necessary to fit the decay properties.
Without inter-doublet interactions ($g_{5,6} = 0$) one can
immediately read off the eigenvalues:
\begin{eqnarray}
M_{\pm}^{(1)} &=& \frac{1}{2}\left[ \sqrt{(g_{1} + g_{2})^2
f_{\pi}^{2} + 4 m_{12}^{2}} \pm (g_{1} - g_{2}) f_{\pi} \right],
\label{e:eigenmass1} \\
M_{\pm}^{(2)} &=& \frac{1}{2}\left[ \sqrt{(g_{3} + g_{4})^2
f_{\pi}^{2} + 4 m_{34}^{2}} \pm (g_{3} - g_{4}) f_{\pi} \right],
\label{e:eigenmass2} \
\end{eqnarray}
where the former two, Eq. (\ref{e:eigenmass1}), correspond to the
first (nucleon) parity doublet and the latter two Eq.
(\ref{e:eigenmass2}), correspond to the second (Roper) parity
doublet. Following Ref.~\cite{jido}, we can determine the coupling
and mass parameters, as well as the mixing angles $\theta_{ij}$,
determined by
\begin{eqnarray}
\tan 2\theta_{ij} &=& \frac{2 m_{ij}}{(g_i + g_j)f_{\pi}} . \
\end{eqnarray}
We show the results in Table \ref{t:coupl1} and Fig. \ref{f:mass}.
\begin{table}[tbh]
\begin{center}
\caption{Coupling constants obtained from the nucleon masses with
doublets ($N(940),N^{*}(1535)$), ($R(1440),N^{*}(1650)$) and the
decay widths $N^{*}(1535) \to \pi N(940)$ and $N^{*}(1650) \to \pi
R(1440)$.(Scenario I)}\label{t:coupl1}
\begin{tabular}{ll}
\hline constant & value \\
\hline
$g_{1}$ &  10.4 \\
$g_{2}$ &  16.8 \\
$m_{12}$ & 270 MeV \\
$\theta_{12}$ & 6.3$^o$ \\
$g_{3}$ &  14.6 \\
$g_{4}$ &  16.8 \\
$m_{34}$ & 503 MeV \\
$\theta_{34}$ & 9.5$^o$ \\
\hline
\end{tabular}
\end{center}
\end{table}
\begin{figure}[tbp]
\centerline{\includegraphics[width=4in,,keepaspectratio]{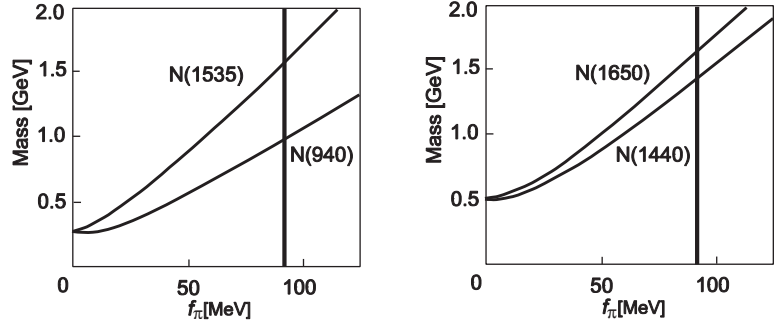}}
\caption{The nucleon masses as functions of $\langle \sigma
\rangle_0 = f_{\pi}$. } \label{f:mass}
\end{figure}

\item {\bf Scenario II}

Once again, the nucleon mass matrix is in a simple block-diagonal
form when the nucleon fields form the following 1$\times$4
row/column "vector": $(\Psi,\Phi) = (N_{m}^{+}, N_{n}^{\prime -},
N_{m}^{\prime +}, N_{n}^{-}) \to (N_{m}^{+}, \gamma_5
N_{n}^{\prime -}, N_{m}^{\prime +}, \gamma_5 N_{n}^{-})$.
\begin{eqnarray}
\lim_{\rm U_A(1) symm.} M &=& \left(\begin{array}{cccc}
g_{1} f_{\pi} & m_{12} \gamma_{5} & g_{5} f_{\pi} & 0 \\
m_{12} \gamma_{5} & g_{2} f_{\pi} & 0 & g_{6} f_{\pi} \\
g_{5} f_{\pi} & 0 & g_{3} f_{\pi} & m_{34} \gamma_{5} \\
0 & g_{6} f_{\pi} & m_{34} \gamma_{5} & g_{4} f_{\pi} \\
\end{array}\right) .
\end{eqnarray}
Note that only the inter-parity-doublet interactions $g_{5,6} \neq
0$ mix these two parity-doublets. We may repeat our analysis as in
the first case, but the data is insufficient to determine all the
couplings as in the Scenario I. We show the results in Table
\ref{t:coupl2}.
Note that here we do not attempt to evaluate the coupling
constants $g_3$ and $g_4$, because the decay of $N(1535)$ to
$R(1440)\pi$ is kinematically forbidden when using the central
values of the resonance masses.

\begin{table}
\begin{center}
\caption{Coupling constants obtained from the nucleon masses with
doublets ($N(940),N^{*}(1650)$), ($R(1440),N^{*}(1535)$) and the
decay widths $N^{*}(1650) \to \pi N(940)$ and $N^{*}(1535) \to \pi
R(1440)$.(Scenario II) }\label{t:coupl2}
\begin{tabular}{ll}
\hline constant & value  \\
\hline
$g_{1}$ &  10.5 \\
$g_{2}$ &  18.1 \\
$m_{12}$ & 295 MeV \\
$\theta_{12}$ & 6.8$^o$ \\
$g_{3}$ &  -- \\
$g_{4}$ &  -- \\
$m_{34}$ & -- \\
$\theta_{34}$ & -- \\
\hline
\end{tabular}
\end{center}
\end{table}
\end{itemize}

Next, we remember that the Roper mixes with the ground state as
well, if it is a non-Abelian mirror field, as in scenario II. In
other words, one must take into account the equal-parity
naive-mirror mixing due to $g_5 \neq 0$. We show the results in
Table \ref{t:coupl3}. Here
\begin{eqnarray}
\tan 2\theta_{13} &=& \frac{- 2 g_{5}}{(g_1 - g_3)} . \
\end{eqnarray}
\begin{table}
\begin{center}
\caption{Coupling constants obtained from the nucleon masses of
the equal-parity doublet ($N(940),R(1440)$).} \label{t:coupl3}
\begin{tabular}{ll}
\hline constant & value  \\
\hline
$g_{1}$ &  15.15 \\
$g_{3}$ &  10.45 \\
$g_{5}$ &  -1.3 \\
$\theta_{13}$ & 14.5$^o$ \\
\hline
\end{tabular}
\end{center}
\end{table}
The predicted coupling strength $g_{\pi N R} = \frac12 (g_{3}-
g_{1}) \sin 2 \theta_{13}$ is significantly smaller than the one
obtained from the decay width $R(1440) \to \pi N(940)$. The order
of magnitude of $g_{1}$ is the same in both fits, so we may
conclude that a simultaneous diagonalization of the complete mass
matrix may lead to complete agreement.

Manifestly, the good $U_{A}(1)$ symmetry limit is sufficient to
reproduce the nucleon spectrum in either scenario. Thence our {\bf
main conclusion}: {\it mass degeneracy of opposite-parity nucleon
resonances is not a consequence of the explicit $U_{A}(1)$
symmetry (non) breaking}. This conclusion was also reached by
Christos, albeit for just one parity doublet and without mirror
fields. In general one has four (quadratic) equations with at
least eight unknowns (six coupling constants and two bare
``masses"). Clearly one needs other input, e.g. the decay widths,
to fix all six parameters. There are too few measured/able decay
widths to fix this ambiguity, however. A complete solution of this
problem is beyond the realm of this paper, anyway.

\subsection{The axial couplings}
\label{s:axial}

The mixing of naive and mirror nucleons leads to a change of the
axial coupling constants, both isovector $g_{A}^{(1)}$ and
isoscalar $g_{A}^{(0)}$. The simple mixing due to the ``mirror
mass" term $m_{12}$ can only reduce the absolute value of both
axial coupling constants from unity, in both scenarios. The mixing
angles $\theta_{ij}$ are shown in Tables \ref{t:coupl1} and
\ref{t:coupl3}, which leads to
\begin{eqnarray}
g_{A}^{(1)} &=& \cos^2 \theta_{12} - \sin^2 \theta_{12}
= 0.976 ,\\
g_{A}^{(0)} &=& \sin^2 \theta_{12} - \cos^2 \theta_{12} = - 0.976,
\end{eqnarray}
in Scenario I, and
\begin{eqnarray}
g_{A}^{(1)} &=& \cos^2 \theta_{12} - \sin^2 \theta_{12} = 0.972 ,\\
g_{A}^{(0)} &=& \sin^2 \theta_{12} - \cos^2 \theta_{12} =
- 0.972,
\end{eqnarray}
in Scenario II. Manifestly, neither of these two values of the
isoscalar axial coupling constant is anywhere close to the
measured one, $g_A^{(0)} = 0.28 \pm 0.16$.

On the other hand, the nucleon-Roper mixing angle $\theta_{13}$,
due to the off-diagonal coupling $g_{5}$ in Scenario II changes
the value of the isoscalar $g_{A}^{(0)}$ to somewhere between its
``Ioffe" value of $-1$ and -3:
\begin{eqnarray}
g_{A}^{(1)} &=& \cos^2 \theta_{13} - \sin^2 \theta_{13} = \cos
2\theta_{13} = 0.875 ,\\
g_{A}^{(0)} &=& -3 \sin^2 \theta_{13} - \cos^2 \theta_{13} = - 2 +
\cos 2\theta_{13} = - 1.125,
\end{eqnarray}
Unfortunately we have already seen that the analogous mixing angle
$\theta_{14}$ in Scenario I can not be determined from the present
analysis, but whatever its value, it would only improve the value
of
\begin{eqnarray}
g_{A}^{(0)} &=& 3 \sin^2 \theta_{14} - \cos^2 \theta_{14} = 1 - 2
\cos 2\theta_{14} \geq -1 .
\end{eqnarray}
We summarize the situation in Table \ref{t:axial}.
\begin{table}
\begin{center}
\caption{Axial coupling constants obtained in different
scenarios.} \label{t:axial}
\begin{tabular}{llll}
\hline constant & ~~~~~~I & ~~~~~II & ~~~~II.A \\
\hline $g_{A}^{(1)}$ & ~~0.976 &  ~~0.972 & ~~0.875 \\
$g_{A}^{(0)}$ & - 0.976 &  - 0.972 &  - 1.125 \\
\hline
\end{tabular}
\end{center}
\end{table}
We are forced to conclude that none of these scenarios lead to a
viable picture of the nucleon ground state (though perhaps some
may be viable for the resonances). Of course, we have not included
the mixing with the $(1,\frac12)\oplus (\frac12,1)$ chiral
multiplet, as yet, which was assumed by Weinberg \cite{weinberg}
to be vital for the isovector axial coupling, and may yet solve
the isoscalar axial coupling problem, as well \cite{NHD}.

\section{Summary and Discussion}
\label{D}

We have analyzed the role of chiral symmetry in general and of the
$U_{A}(1)$ symmetry in particular in the
nucleon-Roper-two-odd-parity-nucleon-resonances system, under the
assumption that the above four nucleon states form a particular
set of chiral multiplets, as implied by the three-quark
construction of the baryon interpolating fields.
The four nucleon fields naturally split into two ``parity
doublets" due to their $U_{A}(1)$ symmetry transformation
properties. We classify the meson-nucleon interactions according
to their $U_{A}(1)$ symmetry transformation properties. It is
crucial to keep all  $U_{A}(1)$ symmetry conserving interaction
terms, even the ``cubic" ones, which are sometimes redundant for
the purpose of mass determination. Yet, note, that {\it if we had
only (incorrectly) insisted on interactions that are linear in
meson fields, we would have been led to the 
different conclusion that the masses of the ``triple Abelian"
nucleon parity doublet are degenerate in the good $U_{A}(1)$
symmetry limit}.

The nucleon mass spectrum and the one-pion decay properties have
been used to fix some of the free coupling constants in the
($\sigma,\pi$) sector, see Tables \ref{t:coupl1}- \ref{t:coupl3}.
Only two $N^{*} \to \eta + N$ decays are kinematically allowed, so
they are not sufficient to determine (all of) the remaining
(${\vec \sigma}, \eta$) couplings.

The insight that the nucleon and the Roper fields may form two
different representations of the $U_A(1)$ symmetry, and that
their mass differences can be explained only by the spontaneous
breaking of $SU_L(2) \times SU_R(2)$ and $U_A(1)$ symmetries, while
explicitly preserving the $U_A(1)$ symmetry, is the main result of
the present paper. A corollary of this result is that the
parity-doublet mass splittings are not determined by the
$U_{A}(1)$ symmetry breaking, as was conjectured in the literature
\cite{jps}. Moreover, the nucleon-Roper mass difference in some
calculations, such as the one of Ref.~\cite{Nagata} in the NJL
model, are not a consequence of the broken $U_{A}(1)$ symmetry in
that model, either.

$U_{A}(1)$ symmetry in nucleon spectra has been discussed before,
most notably by Christos \cite{Christos}, who used only one parity
doublet ($N$(940) and $N^{*}$(1535)), however, and drew
conclusions that are consistent with, but only a small subset of
ours. He argued that the parity doublet mass difference is
proportional to a particular $\eta NN^{*}$ coupling constant,
which is in close agreement with our results. He did not try to
connect other mass differences, such as the Roper-nucleon one, to
this mechanism, as he did not use alternative (``mirror") sets of
fields, which is a novel contribution of our paper (for a
comparison of our formalism with Christos', see 
\ref{app:Christos}).

Jido, Oka and one of us (A.H.) \cite{jido98} studied QCD sum rules
for the odd-parity nucleon resonance $N^{*}$(1535) as a function
of the field and its $U_{A}(1)$ transformation property. We found
that this transformation property is the crucial ingredient
determining the $\eta NN^{*}$ coupling constant. This was perhaps
the first explicit demonstration of the $U_{A}(1)$ symmetry's role
in the odd-parity nucleon spectra; Jaffe {\it et al.} have argued
for the same goal, but along different, more general lines
\cite{jps}: It is well known that spontaneously broken symmetry,
like the $SU_{L}(2) \times SU_{R}(2)$ one, cannot lead to
mass splitting predictions without additional assumptions
\cite{wein66,jps}. $U_{A}(1)$ symmetry is different in this regard
as it is explicitly broken, and badly at that. Different operators
break this symmetry in different ways and this difference might
show up in the mass spectra. An explicitly broken linear Abelian
chiral symmetry such as the $U_{A}(1)$ one, can predict mass
relations in certain special situations, however, as was shown in
the case of scalar mesons in Ref. \cite{vd96}. In the
nucleon-meson problem, however, there is a sufficiently large
latitude to allow a fit of the four nucleon masses and of the most
important nucleon decays without having to invoke the explicit
$U_{A}(1)$ symmetry breaking.

\section*{Acknowledgments}
\label{ack}

We wish to thank Dr. N. Ishii and Prof. W. Bentz, for valuable
conversations about the Fierz transformation of nucleon fields and
(in)dependence of nucleon fields. One of us (V.D.) thanks Prof. H.
Toki for hospitality at RCNP, where this work was started, on two
occasions and Prof. S. Fajfer for warm hospitality at the
Institute Jo\v zef Stefan in Ljubljana where this work was
finished.

\appendix

\section{Comparison with Christos, Ref. [**]}
\label{app:Christos}

Let us consider the two baryon operators,
\begin{eqnarray}
 B^1&=&(q^T_{a} C\gamma^5 i\tau_2 q_{b}) q_c \eps_{abc},\\
 B^2&=&(q^T_{a} C\gamma^5 i\tau_2 q_{b})\gamma_5 q_c \eps_{abc}.
\label{eq:operator1}
\end{eqnarray}
Here indices $a,\;b$ and $c$ label the color of the three quarks,
whereas the Pauli matrices $\tau_i$ operate in the flavor space.
We shall omit the color index from now on. The choice of the
second operator is different from the original choice by Christos
\cite{Christos}, who employed
\begin{equation}
B^3=(q^T C i\tau_2 q )q.
\end{equation}
(In the original paper, $B^3$ was labelled $B^2$ and {\it vice
versa}.) Spin and parity of $B^2$ and $B^3$ are the same.

Christos \cite{Christos} considered the decomposition of $B^{1,3}$
fields in the $L-R$ representation. The operators $B^1$ and $B^3$
are reduced to
\begin{eqnarray*}
B^1_L&=&[-q_L^T C i\tau_2 q_L + q_R^T C i\tau_2 q_R] q_L\\
B^1_R&=&[-q_L^T C i\tau_2 q_L + q_R^T C i\tau_2 q_R] q_R\\
B^3_L&=&[q_L^T C i\tau_2 q_L + q_R^T C i\tau_2 q_R] q_L\\
B^3_R&=&[q_L^T C i\tau_2 q_L + q_R^T C i\tau_2 q_R] q_R
\end{eqnarray*}
We see that in the $L-R$ representation, there are four
independent  operators,
\begin{equation}
[q_L^T C i\tau_2 q_L]q_L,\; [q_R^T C i\tau_2 q_R] q_L,\; [q_L^T C
i\tau_2 q_L]q_R,\; [q_R^T C i\tau_2 q_R] q_R. \label{eq:operator2}
\end{equation}

It is important that the four operators $B^{1,3}_{L,R}$ are
independent (different) as expressed by the four combinations of
the four operators in $L-R$ representation. In other words, we may
rewrite the four independent operators in the $L-R$ representation
by way of $B^{1,3}_{L,R}$. For example,
\begin{equation}
[q_L^T C i\tau_2 q_L] q_L=(B^3_L-B^1_L)/2.
\end{equation}
All four operators of Eq.~(\ref{eq:operator2}) can be expressed in
terms of $B^{1,3}_{L,R}$, none of which are identical. Christos
constructed the chiral Lagrangian by using the operators $B_1$ and
$B_3$.

Next we consider the same algebra with operators
Eq.~(\ref{eq:operator1}),
\begin{eqnarray}
B^1_L&=&[-q_L^T C i\tau_2 q_L+q_R^T C i\tau_2 q_R] q_L\\
B^1_R&=&[-q_L^T C i\tau_2 q_L+q_R^T C i\tau_2 q_R] q_R\\
B^2_L&=&-[-q_L^T C i\tau_2 q_L+q_R^T C i\tau_2 q_R] q_L\\
B^2_R&=&[-q_L^T C i\tau_2 q_L+q_R^T C i\tau_2 q_R] q_R
\end{eqnarray}
In this case we find $B^1_R = B^2_R$ and $B^1_L = - B^2_L$. Thus,
we can not represent four operators in $L-R$ representation with
$B^{1,2}_{L,R}$ independently. Hence, we can not apply Christos'
method to operators $B_1$ and $B_2$.

This tells us that only four helicity components are independent:
a pair of L- and R-handed baryon operators each. As each massive
Dirac field requires two chiral components we conclude that there
are (only) two independent nucleon operators, of either parity.
The parity of the field can be arbitrarily chosen and the opposite
parity field (the parity partner) is necessarily degenerate.


\section{Higher order meson terms in the interaction Lagrangian}
\label{app:cubic}

We consider higher order terms with containing two- and
three-meson fields. The linear meson terms was defined in
Eq.(\ref{ABmeson}) as,
\begin{eqnarray*}
A &=& \big[ \sigma + i \gamma_{5} \mbox{\boldmath$\tau$} \cdot
\mbox{\boldmath$\pi$} \big],\\
B &=& \big[\mbox{\boldmath$\tau$} \cdot \mbox{\boldmath$\sigma$} +
i \gamma_{5} \eta \big].
\end{eqnarray*}
Under $SU(2)_A$ transformation, these terms transform into
\begin{eqnarray}
A\to U^\dagger A U^\dagger,\; B\to U^\dagger B U^\dagger.
\end{eqnarray}
Both of them are covariant under $SU(2)_A$. In contrast they are
not covariant under $U(1)_A$, however we can construct covariant
terms by choosing the linear combinations of them as,
\begin{eqnarray}
\delta_5 (A+B) \to -2i\gamma_5 a (A+B),\\
\delta_5 (A-B) \to +2i\gamma_5 a (A-B).
\end{eqnarray}
Higher power terms are constructed by these linear terms. The term
with two-meson fields are given as
\begin{eqnarray}
M_2 = A A^\dagger - B B^\dagger + A B^\dagger -B A^\dagger,\\
M_3 = A A^\dagger - B B^\dagger - A B^\dagger +B A^\dagger,
\label{eq:msqr}
\end{eqnarray}
These terms are reexpressed by using the determinant term with $M
=\bar{q}_i (1-\gamma_5) q_j$, as
\begin{eqnarray}
A A^\dagger - B B^\dagger &=& \frac{1}{2}(\det M + \det M^\dagger),\\
A^\dagger B - B^\dagger A &=& \frac{1}{2} \gamma_5 (\det M - \det
M^\dagger),
\end{eqnarray}
or explicitly,
\begin{eqnarray}
A A^\dagger - B B^\dagger &=& \sigma^2 + \pi^2 - \eta^2 - \vec{\sigma}^2,\\
A^\dagger B - B^\dagger A &=& 2 i \gamma_5 (\sigma\eta -
{\vec{\sigma}} \cdot {\vec \pi}).
\end{eqnarray}
The transformation under $SU(2)_V\times SU(2)_A$ are
\begin{eqnarray}
M_i\to U^\dagger M_i U,\; U=e^{i\inner{\tau}{a} \gamma_5},
\end{eqnarray}
where $i=2,\;3$.  Note that the difference between the
transformations $M_i$ and $A,\; B$, by which the non-Abelian
naive-mirror mixing terms can be constructed. Under $U_A(1)$
transformation, these terms transform into
\begin{eqnarray}
\delta_5 M_2 &=& 4 i \gamma_5 a M_2,\\
\delta_5 M_3 &=& -4 i \gamma_5 a M_3.
\end{eqnarray}
Finally, we construct the cubic meson terms by the combinations of
the linear and square terms. The $SU(2)_L\times SU(2)_R$ symmetry
allows the following terms to be covariant,
\begin{eqnarray}
&& M_i (A\pm B) \to U^\dagger M_i (A\pm B) U^\dagger, \nonumber \\
&& (A^\dagger\pm B^\dagger) M_i \to U (A^\dagger \pm B^\dagger)
M_i U. \label{eq:trsMAB}
\end{eqnarray}
Of course their hermite conjugates can also be used. However there
is a relation $M_2^\dagger = M_3$ and $M_3^\dagger = M_2$. Then,
the hermite conjugate of the second of Eqs.(\ref{eq:trsMAB}) gives
just the former one. Hence there are eight cubic meson terms.
Finally, we summarize the $U_A(1)$ transformation properties of
the cubic terms.
\begin{eqnarray}
\delta_5 M_2 (A+B) &\to& 2i\gamma_5 a M_2 (A+B),\\
\delta_5 M_3 (A+B) &\to& -6i\gamma_5 a M_3 (A+B),\\
\delta_5 M_2 (A-B) &\to& 6i\gamma_5 a M_2 (A-B),\\
\delta_5 M_3 (A-B) &\to& -2i\gamma_5 a M_3 (A-B),\\
\delta_5 (A^\dagger +B^\dagger) M_2 &\to& 6 i\gamma_5  a(A^\dagger + B^\dagger) M_2,\\
\delta_5 (A^\dagger +B^\dagger) M_3 &\to& -2 i\gamma_5  a(A^\dagger + B^\dagger) M_3,\\
\delta_5 (A^\dagger -B^\dagger) M_2 &\to& 2 i\gamma_5  a(A^\dagger - B^\dagger) M_2,\\
\delta_5 (A^\dagger -B^\dagger) M_3 &\to& -6 i\gamma_5 a
(A^\dagger - B^\dagger) M_3 .
\end{eqnarray}

\eject


\end{document}